\documentclass[11pt,a4paper]{JHEP3}
\usepackage{amssymb}
\usepackage{epsfig,amssymb,euscript}

\def\d{\partial}
\def\1{{\bf 1}}
\def\x{{\bf x}}

\def\d{\partial}

\def\ln{{\rm ln}}
\def\a{\alpha}

\def\0{\nonumber}

\newcommand{\B}{{\mathcal B}}

\newcommand{\N}{{\mathcal N}}

\newcommand\ES{\EuScript{S}}
\newcommand\EC{\EuScript{C}}

\newcommand\ET{\EuScript{T}}

\newcommand\EB{\EuScript{B}}

\newcommand\ER{\EuScript{R}}
\newcommand\EN{\EuScript{N}}

\newcommand\I{\mathbb{I}}
\def\be{\begin{equation}}
\def\ee{\end{equation}}
\def\eea{\end{eqnarray}}     
\def\bea{\begin{eqnarray}}

\font\doppio=msbm10 at11pt
\newcommand{\RR}{\hbox {\doppio R}}

\preprint{hep-th/0311}

\title{Interpolating State in String Field Theory}

\author { D. Mamone \\
International School for Advanced Studies (SISSA/ISAS)\\
Via Beirut 2--4, 34014 Trieste, Italy, and INFN, Sezione di
Trieste\\
E-mail:   \email{mamone@sissa.it}}

\abstract{We derive an oscillator form for the Butterflies in terms of Sliver matrix $S$ and its twisted version $T$ as was already done for the Wedges in term of $T$. We write a General Squeezed state depending on a matrix $U$ and we show in a compact way the interpolation between Identity state and the Sliver and between the Nothing state and the Sliver, growing in powers of $T$ and $S$ matrices, respectively, in the choice of such matrix $U$. Furthermore, we define a class of states which we call Laguerre states and we give a formal derivation of such interpolating state in terms of them.}


\begin{document}

\section{Introduction}

In Witten's Open String Field Theory \cite{W1}, the information about open strings interaction is encoded in an associative but noncommutative multiplication between Open String Fields. This multiplication is the Witten's star product and defines an algebra.
The main problem in Witten's Open SFT is that we don't know how to solve the equation of motion. Rastelli, Sen and Zwiebach \cite{RSZ1,RSZ2} proposed a new formulation of SFT, conjectured to be the SFT written at the ``true'' vacuum, the minimum of the tachyon potential. This formulation is know as Vacuum String Field Theory. The advantage of this theory is that we are able to find exact solutions: D--branes are seen to correspond to open string fields that are projectors of the star algebra \cite{KP,RSZ2}. This happens because the Witten's kinetic operator which contains both matter and ghost operators, is replaced by a purely ghost one and this simplifies the equation of motion, which for the matter sector becomes just a projector condition 

\be
\Psi * \Psi = \Psi
\ee

\noindent under the Witten's star product.
 For this reason, star projectors are important in VSFT. There are several of such states that were found: the identity string field, the ``Sliver'' and the family of so called ``Butterflies'' \cite{GRSZ2}. They can be described in a conformal formalism as Riemann surface whose boundary consists of a parametrized open string and a piece with boundary conditions. Such states are called surface states and are defined by a conformal map, the configuration of the open string. In oscillator formalism, using creation and annihilation operators of the BCFT, these states have a squeezed form involving infinite dimensional matrices.
The Sliver, for instance, has the form

\be 
|\Xi\rangle = \exp \left(  -\frac{1}{2} \sum_{m,n=1}^\infty \left(S\right)_{mn}a^\dagger_m
a^\dagger_n \right) |0 \rangle \,.
\ee

\noindent where $S$ is a matrix which can be expressed in terms of the Neumann coefficients of the interaction vertex of SFT \cite{leclair1,leclair2,GJ1,GJ2}or as residues of poles using the corresponding conformal map.
In \cite{RSZ2} Rastelli, Sen and Zwiebach derived the Sliver in an algebraic way using the oscillator representation and showed numerically that this solution is the same found with the conformal formalism and representing the D25 brane. Okuda proved it analitically in \cite{okuda}.\\
 There are two big families of surface states studied in SFT, the so called ``Wedge'' states and the Butterflies.
For Wedge states, the Riemann surface is an angular sector of the unit disk, with the left-half and the right-half of the open string being the two radial segments and the arc having open string boundary condition. Such state is thus defined by the angle at the midpoint of the string and this angle adds under star moltiplication since star product between two string fields can be seen as the joining of the right-half part of the first string with the left-half part of the second one. In this picture, the Identity state and the Sliver, which belong to wedge states, can be interpreted respectevely as the zero and the infinity angle wedge state. Indeed, they are the only projectors among wedge states: the sum of two zero or two infinities angle is still a zero or an infinite angle.
Wedge states form an abelian subalgebra of star algebra.
Butterflies, on the other hand, are all star projectors. They are singular configuration in which the midpoint of the open string is sent to infinity. We will see they depends on a real paramenter $\alpha$. They share only one state with the wedges: the Sliver belongs also to Butterflies. The Sliver can be seen at the same time as the Nth wedge with $N$ going to infinity or as the $\alpha$th butterflies with $\alpha$ going to zero.

The first aim of this paper is to investigate the star algebra and to recognise
in those projectors like Butterflies which were not found with algebraic oscillator formalism, a common structure with the Wedges, studying the crucial difference which makes them projectors. This is done in order to
look for new tools to do calculations in the algebraic operatorial formalism and to guess which could be the form of a good basis on which expand and hopefully solve the equation of motion of the Witten's SFT. 

For this reason we derive an oscillator representation for Butterflies and we
show that both Wedges and Butterflies States can be rewritten in a General squeezed state form  

\be
|G_U\rangle={\N_\Xi \over \left (det\left(1+TU\right)\right )^{1/2}} \, e^{-{1 \over 2}\,CG\,a^\dagger a^\dagger}|0\rangle
\ee

where

\be
G=\frac{U + T}{ {\bf 1}+ TU}\
\ee

depends on a matrix $U$. The interesting point is that if we choose matrix $U$ to be powers of the Sliver $S$ matrix we obtain the family of Butterflies, on the other hand, if we choose to be powers of its twisted version $T$, we get Wedge states, which are not all projectors.
In other words, such state interpolates between Wedges and Butterflies.
In particular we have: 

\begin{itemize}

\item
$U=0$ corresponds to the Sliver state

\item
$U=1$ corresponds to the Identity state

\item
$U=C$ corresponds to the Nothing state

\item
$U=-S$ corresponds to the Butterfly state

\item
$U=\left(-T\right)^{N-1}$ corresponds to the Nth Wedge state

\item
$U=\left(-S\right)^{N-1}$ ($N$ even) corresponds to the $\alpha={2 \over N}$ Generalized Butterfly state

\end{itemize}

The second aim is to suggest a formal way to obtain such General state using a suitable resummation of states that we call ``Laguerre'' states which are a modification of ``Ancestors'' solutions found in \cite{BMS3}. These latter are an infinite class of projectors defined acting with a Laguerre polynomial expression  of string creation operators on the Sliver, whose low-energy limit give rise to GMS solitons. For this reason they were called ``Ancestors''.
In the final part, we will point out that partial isometry symmetry of VSFT seems to suggest that Ancestors solutions behave like a complete basis, maybe for a star subalgebra.\\

The paper is organized as follows. In section 2 we review definitions of surface states following \cite{GRSZ2}. In section 3 we derive an oscillator representation for the Butterflies starting from the Moyal wave-functional one given in \cite{fuchs}, which we will recognise in section 4 in our interpolating state. 
In section 4, we write the General Squeezed State showing various choices for matrix $U$ that give rise to various string fields. For our calculations, we will use the diagonal basis used in \cite{fuchs}, which consists of eigenvectors of the operator $K_1^2$, instead of the usual $K_1=L_1+L_{-1}$ and allows us to rewrite Sliver and twist matrices as two dimensional matrices. This turns out to be very useful for Butterflies representation.\\
In section 5 we define Laguerre states and a function of them that we will identify with the General Squeezed state written in section 4.   

\section{A short review of surface states and star projectors}

In this section we brifly review the definition of surface states through the conformal maps which define them and we will recall the oscillator representation of each of them, which has the form of a squeezed state. 
We mainly refer to \cite{GRSZ2}. 
These surface states are Riemann surfaces whose boundary
consists of a  parametrized open string and a piece with open string boundary conditions. 
Thinking of a surface state as a string field, it is possible to define
a geometric operation which corresponds to star product in SFT:
one glues the right-half of the open string in the first surface to 
the left-half of the open string in the second surface, 
and the surface state corresponding to the glued surface is the 
desired product. Usually, multiplication of a surface state to itself
leads to a surface state that looks different from the initial
state. This is the reason why it is not trivial to find 
projectors.
Let us define the thing more precisely.

\medskip
A surface state $\langle\Sigma$ is a state related with a Riemann surface $\Sigma$ with the topology of a disk $D$, with a marked point $P$, the puncture, lying on the boundary of $D$, and a local coordinate around it. Local coordinate at a puncture is obtained from an analytic map $m$ taking a canonical half-disk $H_U$ defined as
\be
H_U : \{ |\xi| \leq 1, {\rm Im}(\xi) \geq 0  \}
\ee
into $D$, where $\xi = 0$ maps to the puncture $P$, and the image of the real segment $\{|\xi| \leq 1, {\Im}(\xi) = 0 \}$ lies on the boundary of $D$. The coordinate $\xi$ of the half disk is called the local coordinate. Using any global coordinate $u$ on the disk $D$, the map $m$ can be described by some analytic function $f$:
\be
 u= f(\xi), \quad u(P) = f(0)
\ee
Given a BCFT with state space $H$, the state 
$\langle\Sigma| \in H^*$ associated to the surface $\Sigma$ is defined as follows. For any local operator $\phi (\xi)$, with associated state 
$\phi\rangle = \lim_{\xi \to 0} \phi(\xi)|0\rangle$ we set
\be
\langle\Sigma \phi \rangle = \langle f \circ \phi(0)\rangle_D
\ee
where $\langle\,~\rangle_D$ is the correlation function on $D$ and 
$s \circ \phi(0)$ is the transform of the operator by the map $f(\xi)$. 
The gluing of surfaces, conformal analog of star product of string field states, requires a well defined full map of the half disk $H_U$ into the disk $D$.

\subsection{Operator representation of surface states} 

Let us see the representation of surface states in terms of 
Virasoro operators acting on the SL(2,R) invariant vacuum.

We can write the surface state $\langle\Sigma|$ 
as
\be \label{Uf}
\langle \Sigma | = \langle 0| U_f \equiv \langle 0| \exp\left(\sum_{n=2}^\infty
v_n^{(f)} L_n \right) \, ,
\ee
where the coefficients $v_n^{(f)}$ are determined
by the condition that the vector field 
\be \label{edefvxi}
v(\xi) = \sum_{n=2}^\infty v_n^{(f)} \xi^{n+1}\, ,
\ee
exponentiates
to $f$,
\be \label{vdef}
\exp \left( v(\xi) p_\xi \right) \xi
= f(\xi) \,.
\ee
We now
consider the one-parameter family of maps
\be \label{fbeta}
f_{\beta} (\xi) =  \exp \Bigl(\beta \, v(\xi) {\partial\over \partial \xi}
\Bigr)\, \xi
    \,.
\ee
This gives
\be \label{vf}
 \frac{d}{d \beta}  f_\beta(\xi) = v(f_\beta (\xi)) \,.
\ee
Solution, taking into account the boundary condition
$f_{\beta=0}(\xi)=\xi$, gives:
\be \label{egenc1b}
f_\beta(\xi) = g^{-1}(\beta + g(\xi))\, ,
\ee
where
\be \label{egenc2}
g'(\xi) = {1\over v(\xi)}\, .
\ee
Thus
\be \label{egenc1}
f(\xi) = g^{-1}(1 + g(\xi))\, .
\ee
Equations (\ref{egenc2}) and (\ref{egenc1}) give $f(\xi)$ if $v(\xi)$
is known. 

\subsection{Oscillator representation of surface states} 

We consider the matter part of the state and the oscillators will be associated
to free scalar fields of the Boundary CFT.
If $a_m$, $a_m^\dagger$ denote the annihilation and creation operators 
we have:
\be \label{soscill}
|\Sigma\rangle =\exp \left(  -\frac{1}{2} \sum_{m,n=1}^\infty a^\dagger_m
V^f_{mn} a^\dagger_n \right) |0 \rangle \,.
\ee

and

\be  \label{evmn}
V^f_{mn} = {(-1)^{m+n+1}\over \sqrt{mn}}
\oint_0{dw\over 2\pi i}
\oint_0{dz\over 2\pi i}
\,{1\over z^m w^n} {f'(z)
f' (w)\over (f(z) - f(w))^2}\,.
\ee

Both $w$ and $z$ integration contours are circles around the origin,
inside the unit circle and 
with the $w$ contour outside the $z$ contour.

\medskip  
The crucial point is that when the vector field $v(\xi)$, generating the
conformal map $f(\xi)$, is known we can put the integral expression 
for the matrix $V^f$ of Neumann coefficients
in another form.
We consider the matrix $V (\beta)$
associated to the family of maps (\ref{fbeta}),
and rewrite (\ref{evmn}) as
\be
V_{mn}(\beta) \equiv V_{mn}^{f_\beta} =  {(-1)^{m+n+1}\over \sqrt{mn}}
\oint_0{dw\over 2\pi i}
\oint_0{dz\over 2\pi i}
\,{1\over z^m w^n} \frac{\partial}{\partial z} \frac{\partial}{\partial w}
\log(f_\beta(z) - f_\beta(w))\, .
\ee
Taking a derivative with respect to the parameter
$\beta$,

\be
\frac{d}{d \beta}  V_{mn}(\beta) = \frac{(-1)^{m+n+1}}{  \sqrt{mn}}
\oint_0{dw\over 2\pi i}
\oint_0{dz\over 2\pi i}
\,{1\over z^m w^n} \frac{\partial}{\partial z} \frac{\partial}{\partial w}
\left( \frac{  v(f_\beta(z)) - v(f_{\beta}(w))}{f_\beta(z) -
f_\beta(w)} \right)\, ,
\nonumber
\ee

Integrating by parts in $z$ and $w$: 
\be \label{Vbeta}
\frac{d}{d \beta}  V_{mn}(\beta) = {(-1)^{m+n+1}  \sqrt{mn}}
\oint_0{dw\over 2\pi i}
\oint_0{dz\over 2\pi i}
\,{1\over z^{m+1} w^{n+1}}  \frac{  v(f_\beta(z))
- v(f_{\beta}(w))}{f_\beta(z) -
f_\beta(w)} \, .
\ee
Neumann coefficients $V_{mn}(\beta=1)$
can be calculated integrating over $\beta$.  

\medskip
There are two big families of surface states studied in SFT: the Wedge states and the Butterflies. 

\subsection{Wedge states} 

The identity string field state and the so called Sliver, the state obtained star-multiplying an infinite number of vacuum, are particular wedge states. 
In this case, the Riemann surface is an angular sector
of the unit disk, with the left-half and the right-half of the
open string being the two radial segments, and the unit radius arc
having the open string boundary conditions. A wedge state is
thus defined by the angle at the open string midpoint, and the star moltiplication between two wedge states gives rise to the wedge state which corresponds to the sum of the angles of the states. The identity string field
is the wedge state of zero angle, and the sliver is the wedge
state of infinite angle.

They are defined by the map 
\be
w_n = \check{f}_n(\xi) \equiv (h(\xi))^{2/n} = 
\bigg(\frac{1 + i\xi}{1 -i \xi}\bigg)^{2/n}
\label{wedgetrans}
\ee
 that sends the upper half-disk $H_U$ into a wedge with the angle at $w_n = 0$ equal to $2 \pi/n$. The transformation (\ref{wedgetrans}) can be rewritten as 
\be
w_n = \exp\bigg(
      i \frac4n \tan^{-1}(\xi)
\bigg) 
\label{tan1}
\ee

We define $\langle \ET_N$ such that
 
\be
\langle \ET_N | \phi \rangle \equiv \langle \check{f}_n \circ \phi (0)\rangle_{D_0}
\label{wedge1}
\ee

The state that we obtain for $N=1$ is the identity state, that in the coordinates $w_N$ is the full unit disk $D_0$ with a cut on the negative real axis. The left-half and the right-half of the string coincides along this cut. The $n \to \infty$ limit is the Sliver. It is an infinitely thin sliver of the disk $D_0$ around the positive real axis. 

We describe now $|\ET_N\rangle$ taking back the wedge on the upper half plane. We define

\be
z_n = h^{-1}(w_n) = i \frac{1-w_n}{1+w_n} = 
\tan\bigg(-\frac{i}{2}\ln w_n \bigg) 
\label{tan2}
\ee

Putting (\ref{tan1}) and (\ref{tan2}) we have

\be
z_n = \tan\bigg( \frac2n \tan^{-1}(\xi) \bigg) \equiv \tilde f_n (\xi)
\ee

and

\be
\langle \ET_N| \phi \rangle = \langle \tilde f_n \circ \phi(0) \rangle_{D_H}
\label{wedge2}
\ee

The description of the Sliver as the $N \to \infty$ wedge looking at (\ref{wedge1}) and (\ref{wedge2})seems to be singular, in the sense that the maps $\check f_N(\xi)$ and $\tilde f_N (\xi)$ are singular in the $N \to \infty$ limit. This apparent singular behaviour is solved by noticing the SL(2,$\RR$) invariance of the correlation functions on the upper half plane. Given any  SL(2,$\RR$) map $R(z)$ we have the relation

\be
\langle \prod_i \O_i(x_i) \rangle_{D_H} = 
\langle \prod_i R \circ \O_i (x_i) \rangle _{D_H}
\ee

for any set of operators $\O_i$ and with $D_H$ denoting the upper half plane. Since the sliver $|\Xi\rangle$ is defined through a correlation function, we can set

\be
R_N (z) = \frac N2 z
\ee

 so that

\be
\langle\Xi| \phi \rangle = \langle f \circ \phi(0) \rangle_{D_H}
\ee

where

\be \label{tanmap}
f(\xi) =  \lim_{N \to \infty} R_N \circ \tilde f_N (\xi) 
       =  \lim_{N \to \infty} \frac N2 \tan \bigg( \frac 2N \tan^{-1}(\xi)
\bigg) = \tan^{-1} \xi
\ee

Under the map

\be
\widehat w_N =  (w_N)^{N/2}
\ee

the unit disk $D_0$ in the $w_N$-coordinates is mapped to a cone in the $\widehat w_N$ coordinate, subtending an angle $N\pi$ at the origin 
$\widehat w_N = 0$. The disk $D_0$ mapped in this way represents then a wedge 
$|\ET_N$.  We can give now the prescription for the $*$ product. Les us consider directly wedge states and remove the local coordinate patch from the disk 
$D_0$ in the $w_N$ coordinate: the left over region becomes a sector of angle 
$\pi(N-1)$. If we denote by $|\ER_\a\rangle$ a sector state arising from a sector of angle $\a$, we have the identification of sector states with wedge states

\be
|\ET_N\rangle = |\ER_{\pi(N-1)}\rangle
\ee

The operation of $*$ multiplication of two wedge states 
$|\ET_M\rangle* |\ET_N\rangle$ is then given by gluing together the two sector states 
$|\ER_{\pi(M-1)}\rangle$ and $|\ER_{\pi(N-1)}\rangle$ identifying the left-hand side of the string front of $|\ET_m\rangle$ with the right-hand side of the string front of $\ET_n\rangle$. With this prescription we obtain the rule

\be
|\ER_a\rangle \ast  |\ER_b\rangle = |\ER_{a+b}\rangle
\ee

that means

\be
|\ET_M\rangle * |\ET_N\rangle = |\ET_{M+N-1}\rangle
\ee

The sliver state ($N \to \infty$) is a projector under the $*$ product: 

\be
|\ER_\infty\rangle \ast  |\ER_\infty\rangle = |\ER_\infty\rangle
\ee

\bigskip
Let us give the operator representation of wedge states. 
We consider $U = U(f_n)$ depending only on matter Virasoro generators $L_n$ and ghost fields $b$ and $c$ such that $\langle n| = \langle 0|U$. Since a primary field of conformal weight $d$ transforms under finite conformal transformation $f$ as

\be
f \circ \phi(z) = (f'(z))^d \phi (f(z))
\ee

we can write 

\be
(f'(z))^d \phi (f(z)) = U_f \phi (z) U^{-1}_f
\ee

with

\be
U_f = \exp[v_0 L_0] \exp\bigg[\sum_{n\geq 1} v_n L_n\bigg]
\ee

The coefficients $v_n$ can be determined recursively from the Taylor expansion of $f$, by requiring 

\be
e^{v_0}  = f'(0) 
\label{finiteconf} 
\exp\bigg[\sum_{n\geq 1} v_n z^{n+1}\d_z \bigg] =  
(f'(0))^{-1} f(z) = z + a_2 z^2 + a_3 z^3 + \ldots 
\ee

For instance, for the first coefficients one finds 

\be
v_1 = a_2\,,\quad v_2 = -a_2^2 + a_3\,,\quad 
v_3 = \frac32 a_2^2 -\frac52 a_2a_3 + a_4
\ee

One can determine eqs.(\ref{finiteconf}) in the following way. Using the commutation relation

\be
[L_m, \phi_n] = ((d-1)m -n)\phi_{m+n}
\ee

we have

\be
U_f \phi(z) U_f^{-1} = \exp [v(z) \d_z + d v'(z)] \phi(z)
\ee

where the function $v(z)$ is such that 

\be
e^{v(z)\d_z} z = f(z)
\ee
 
Choosing $\tilde f_N (z) = \tan \big(\frac2N \tan^{-1} (z)\big)$ to define the wedge states $|\ET_N\rangle$, we have

\be
|\ET_N\rangle   =  \exp 
               \left[ - \frac{N^2 -4}{3 N^2}L_{-2} + \frac{N^4 -16}{30 N^4}L_{-4}
              -\frac{(N^2 - 4)(176 + 128 N^2 + 11N^4)}{1890 N^6}L_{-6} +\cdots \right ] |0\rangle
\ee

Among them, it is possible to recognise, for particular value of $N$:

the Identity State ($N = 1$) (see \cite{fengidentity}):
\be
|I\rangle = \exp\left[
                  L_{-2} -\frac12 L_{-4} + \frac12 L_{-6} + \cdots  
\right] |0\rangle
\ee

and the Sliver State ($N \to \infty$):
\be
|\Xi\rangle  =   \exp \left[ -\frac13 L_{-2} + \frac{1}{30} L_{-4} -\frac{11}{1890}L_{-6}
             + \cdots
\right] |0\rangle
\ee

\medskip

\subsubsection{Oscillator representation of wedge states}

The oscillator representation of the Sliver as given in \cite{KP,RSZ2} is

\be \label{oscsliver}
|\Xi\rangle = \exp \left(  -\frac{1}{2} \sum_{m,n=1}^\infty \left(CT\right)_{mn}a^\dagger_m
a^\dagger_n \right) |0 \rangle \,.
\ee

where the $T$ matrix was found algebraically and written as

\be 
T= {1\over 2X} (1 + X - \sqrt{(1+3 X)(1 - X)} )\, .
\ee

in terms of the infinite dimensional matrix $X=CV^{11}$ which is known: $V$ re the Neumann coefficients and $C=(-1)^m \delta_{mn}$.
Matrix elements of $S$ match with those calculated in the conformal way

\be  
\widehat S_{mn} = {(-1)^{m+n+1}\over \sqrt{mn}}
\oint_0{dw\over 2\pi i}
\oint_0{dz\over 2\pi i}
\,{1\over z^m w^n} {f'(z)
f' (w)\over (f(z) - f(w))^2}\,.
\ee

with $f(z)=tan^{-1}z$.\\

Similarly, for the Identity state we have

\be \label{oscid}
|I\rangle = \exp \left(  -\frac{1}{2} \sum_{m,n=1}^\infty \left(C\right)_{mn}a^\dagger_m
a^\dagger_n \right) |0 \rangle \,.
\ee

As we already said, the previous state are particular cases of the Nth Wedge state which is
 
\be \label{oscwedge}
|\ET_N\rangle_m = \exp \left(  -\frac{1}{2} \sum_{m,n=1}^\infty \left(CT_N\right)_{mn}a^\dagger_m
a^\dagger_n \right) |0 \rangle \,.
\ee

where $T_N$ is the matrix

\be
T_N={\left(-T\right)^{N-1} +T \over 1 - \left(-T\right)^N}
\ee

This relation between the Nth wedge matrix and the twisted sliver matrix $T$ will be useful in the following.

\subsection{The Butterfly State} 

The Butterfly and the Nothing state are particular Butterflies \cite{GRSZ2}. We already said that Generalized Butterfly depends on a parameter $\alpha$ in a way we will see later on. The Butterfly, the first of the family to be found, is simply the $\alpha=1$ case, while the so called Nothing state corresponds to $\alpha=2$. 

The butterfly state is defined by a map from $\xi$ to the upper half $z$ plane

\be
\label{defbut}
z = {\xi\over \sqrt{1+ \xi^2}}\equiv f_{\EB}(\xi)\, ,
\ee

more precisely the surface state $|\EB\rangle$ that defines the butterfly is such:

\be \label{ebutterdef}
\langle\EB|\phi\rangle = \langle f_\EB\circ\phi(0)\rangle_{UHP}\, .
\ee

In the $\xi$-plane we have the half-disk: the circumference is the string, the point $\xi=i$ the midpoint.

In the $z$-coordinate, the open string $|\xi| =1, \Im (\xi) \ge 0$ is mapped to the hyperbola $x^2-y^2 = {1\over 2}$ with $z= x+ iy$.
The fact that $z(\xi=i) = \infty$ means that the open string midpoint coincides with the boundary of the disk.
If we define

\be
\hat z = sin^{-1}\frac{\xi}{\sqrt{1+\xi^2}}
\ee

and

\be
\hat w = e^{2i\hat z} =  e^{2i sin^{-1}\frac{\xi}{\sqrt{1+\xi^2}} }
\ee

The Riemann surface associated to the Butterfly is the following: the half-disk of the $\xi$-plane is mapped to the right-half of the disk in the $\hat w$-plane. The midpoint is sent to the center of the disk and the open string is the vertical diameter. All the rest of the upper half plane is mapped to the left-half of the disk with a cut along the negative real axis, which represents that fact that the midpoint touchs the boundary. (See figure 1, with $\alpha=1$.)


\subsubsection{Operator representation of the butterfly state} 
\label{sbutterope}

We can represent the butterfly $|\EB_t\rangle$ in the operator 
formalism. 
We will use the ``reguleted'' butterfly $|\EB_t\rangle$ defined by the map

\be  z =  f_t(\xi) =  \frac{\xi}{\sqrt{1+ t^2 \xi^2}}=
\exp \Bigl( v_t (\xi) {\partial\over \partial \xi}
\Bigr)\, \xi \,.
\ee

The regulator parameter $t$ must therefore satisfy $t<1$. (See \cite{GRSZ2} for details). We recover the butterfly when $t=1$. \\

Eqs. (\ref{egenc1}), (\ref{egenc2}) give

\be \label{vbutter}
v_t(\xi) = -t^2  \, \xi^3 /2 \, .
\ee

Eq. (\ref{Uf}), (\ref{vdef}) now gives:
\be \label{Bexp}
|\EB_{\,t} \rangle = \exp \left( -\frac{t^2}{2} \,L_{-2} \right)|0\rangle\,.
\ee

\subsubsection{Oscillator representation of the butterfly state} 

Let us represent the matter part of the regulated butterfly state in 
the oscillator representation.
Take $\beta \equiv t^2$,

\be
v(\xi) = -\frac{\xi^3}{2}  \, ,\qquad
f_{\beta} (\xi)= \frac{\xi}{\sqrt{1 + \beta \xi^2}} \,.
\ee

Eq. (\ref{Vbeta}) gives

\begin{eqnarray} 
\frac{d}{d \beta}  V^B_{mn}(\beta) = (-1)^{m+n} \frac{ \sqrt{mn}}{2}
\oint_0{dw\over 2\pi i}
\oint_0{dz\over 2\pi i}
\,{1\over z^{m+1} w^{n+1}}  \frac{ f_\beta(z)^3
- f_\beta(w)^3}{f_\beta(z)-f_\beta(w)}\\
= (-1)^{m+n} \frac{ \sqrt{mn} } {2}
\oint_0{dw\over 2\pi i}
\frac{f_\beta(w)}{w^{m+1}}
\oint_0{dz\over 2\pi i}  \frac{f_\beta(z)}{z^{m+1}}
= (-1)^{m+n} \frac{ \sqrt{mn}}{2}\,  x_m x_n \, , \nonumber
\end{eqnarray}

where

\be
x_m = \oint_0{dw\over 2\pi i}  \frac{f_\beta(w)}{w^{m+1}}
=(-\beta)^{\frac{m-1}{2}}
\frac{\Gamma[\frac{m}{2}] } {\sqrt{\pi}
\Gamma[ \frac{m+1}{2} ]}\quad {\rm for } \; m \;{\rm odd}\,,
\ee

\be
x_m= 0  \quad \qquad \qquad \qquad \qquad {\rm for }\;
m\; {\rm even}\,. \nonumber
\ee

Integrating (\ref{Vbetabutt}) with the initial condition $V(\beta=0)=0$, we
find the Neumann coefficients of the regulated butterfly $(\beta \to 
t^2)$:

\be \label{Vbutt}
V^B_{mn} (t) =  -(-1)^{\frac{m+n}{2}} \frac{\sqrt{mn}} {m+n}
\frac{\Gamma[\frac{m}{2}]
\Gamma[\frac{n}{2} ] } {\pi \Gamma[ \frac{m+1}{2} ]
\Gamma[\frac{n+1}{2}]}\, t^{m+n}\,,
\quad {\rm for } \; m  \; {\rm and}
\;n \;{\rm odd}\,,
\ee

\be
V^B_{mn} (t) =  0\,, \qquad \qquad \qquad \qquad \qquad \qquad \qquad \quad \quad 
\;\;\quad
{\rm for } \; m  \; {\rm or}
\;n \;{\rm even}\,. \nonumber
\ee

\subsection{The Nothing State}

The nothing state is defined by the map:

\be \label{enoth2}
f_{\EN}(\xi) = \frac{\xi}{\xi^2 + 1}
\ee

$V^f_{mn}$ computed using (\ref{evmn}), (\ref{enoth2}) turns out to be equal 
to $\delta_{mn}$. Thus
the oscillator representation of the matter part of the nothing 
state is given by:

\be \label{enoth4}
|\EN\rangle = \exp \left(  -\frac{1}{2} \sum_{m,n=1}^\infty a^\dagger_n
a^\dagger_n \right) |0 \rangle \,.
\ee

\subsection{Generalized butterflies}

The generalized butterfly state $|\EB_g (\alpha)\rangle$ is defined
through a generalization of eq. (\ref{defbut}) to

\be \label{e2}
z= {1\over \alpha} \sin(\alpha \tan^{-1}\xi)\equiv 
f_\alpha(\xi)\, .  
\ee

Note that the case $\alpha=1$:
\be \label{efa1}
f^{\EB_g}_{\alpha=1} = {\xi\over \sqrt{1+\xi^2} }\, .
\ee
so $|\EB_g (\alpha=1)\rangle$ is nothing but the butterfly while 
for $\alpha=0$ we have :
\be \label{efa}
f^{\EB_g}_{\alpha=0} = \tan^{-1}\xi\, .
\ee
so $|\EB_g (\alpha=0)\rangle$ is the
sliver. \\
$|\EB_g(\alpha)\rangle$ is a family of
projectors, interpolating
between the butterfly and the sliver.
For $\alpha=2$ we have the map
\be \label{efa2}
f^{\EB_g}_{\alpha=2} = {\xi \over 1 + \xi^2}\, .
\ee
which corresponds to the `nothing' state.

We can regularize the
singularity at the midpoint and define the
regularized butterfly
by generalizing \ref{defrbut} to
\be \label{e6}
z=f_{\EB_g,t}(\xi) =
{1\over \alpha} { \tan(\alpha\tan^{-1}\xi) \over
\sqrt{1 + t^2 \tan^2 (\alpha\tan^{-1}\xi)} }\, .
\ee
In the $\widehat z$ plane we get
\be \label{e7}
\langle \EB_g(\alpha,t)|\phi\rangle = \langle f^{(0)}\circ
\phi(0)\rangle_{
\EC_{\alpha,t}}\, ,
\ee
where $\EC_{\alpha,t}$ is the image of the upper half $z$ plane in the
$\widehat z$ coordinate system and $f^{(0)}(\xi)=\tan^{-1}\xi$.

For $\alpha=2$ the region of $\EC_{\alpha,t}$ collapses to nothing.
For this  reason the
associated surface state is called the ``Nothing'' state.

\begin{figure}[!ht]
\leavevmode
\begin{center}
\epsfysize=6cm   
\epsfbox{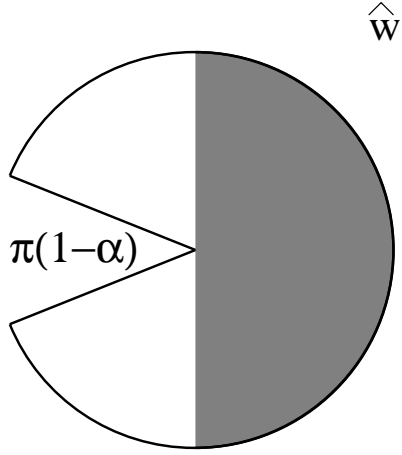}
\end{center}
\caption{The image of
$\EB_g(\alpha)$ in the complex $\widehat w=e^{2i\widehat z}$ plane.
The shaded region denotes the
local coordinate patch. The vertical diameter is the open string and the center of the disk is the midpoint which is sent to infinity. All the rest of HUP is mapped in the white region.} \label{f2}
\end{figure}

\section{Oscillator representation of Butterflies}

In this section 

We will derive it from the Moyal coordinate wave-function representation due \cite{bars,doug,belov}. Let us make first of all, an example with the sliver state.
It is very easy to obtain the gaussian form of the Sliver in the $x(k),y(k)$ Moyal coordinate:

\be
\Psi_\Xi(x(k),y(k))=\langle x(k),y(k)|\Xi\rangle=2e^{-\frac{x^2(k)+y^2(k)}{\theta}}
\ee

we must take the braket between the sliver state in diagonal form \cite{Chen}

\be
|\Xi\rangle =\N_\Xi e^{-\frac{1}{2}\int_0^\infty dks(k)\left( e^{\dagger^2}+o^{\dagger^2}\right)}|0\rangle
\ee

and the Moyal coordinate squeezed state (the same for $y(k)$)

\be
\langle x(k)|=\langle 0|exp\left(e^2+i\sqrt{2}ex(k)-1/2x^2(k)\right)
\ee

In order to do this, we use the formula 

{\small
\begin{eqnarray}
 \label{eten1}
 \langle 0| \exp\Big (-{1\over 2}  a\cdot  P a +
\lambda\cdot a \Big) 
\exp\Big (-{1\over 2} a\cdot Q a^{\dagger} + 
\mu \cdot  a\Big) 
|0\rangle  \\
= \det(1 - PQ)^{-1} \times \\ \times
\exp\Big(\mu^T \cdot (1-PQ)^{-1} \cdot \lambda
-{1\over 2} \mu^T \cdot Q (1-PQ)^{-1} \cdot \mu
 -{1\over 2} \lambda^T \cdot P (1-PQ)^{-1} \cdot
\lambda\Big)\, ,\nonumber
\end{eqnarray}
}

with

\be
\lambda=i\sqrt{2}x(k) \ , \ P=1 \ , \ \mu=0 \ , \ Q=s(k)
\ee

Now, if we would like to do the opposite, starting from the Moyal function representation and get the oscillator one, we should start from the coefficient $p$ of $x(k)$ (or $y(k)$, for the sliver is the same) and, after some algebra, obtain the coefficient $Q$ from

\be
{Q\over 1+Q}-{1\over2} = p
\ee

In our case

\be
p=-{1\over\theta}
\ee

\be
\theta(\kappa)=2\tanh\left(\frac{\kappa\pi}{4}\right)\,,
\ee

then

\be
Q={\theta-2 \over \theta+2}=s(k)
\ee

which is the well-known coefficient of the diagonal form of the sliver \cite{Chen}.

Starting from the Moyal function form of the butterfly, \cite{fuchs}, the same game leads to

\be
{Q_x\over 1+Q_x}-{1\over2} = p_x
\ee

\be
{Q_y\over 1+Q_y}-{1\over2} = p_y
\ee

with 

\be
p_x=-{1 \over 2} , p_y=-{2\over\theta^2}
\ee

thus

\be
Q_x=0 ,  Q_y={\theta^2-4 \over \theta^2+4}={2s \over 1+s^2}
\ee

and we obtain

\be
|\B\rangle=\N_{\B} \, e^{-{1\over2} \, {2s \over 1+s^2} \, o^{\dagger^2}}|0\rangle
\ee

which satisfies the fact simple Butterfly is twist odd.

Moyal wave functions in this diagonal coordinate space $\vec X_\kappa \equiv (x_\kappa,y_\kappa)$
are proportional to the Gaussian

\begin{equation}
\exp\left(-\frac{1}{2}\int_0^\infty
    \vec X_\kappa \tilde L_\kappa \vec X_\kappa d\kappa\right)\,.
\end{equation}

The generalized butterfly $|\B_g\rangle$ in this representation is \cite{fuchs}

\be
\tilde L_\kappa=\coth\left(\frac{\kappa \pi}{4}\right)
    \pmatrix{\tanh(\frac{\kappa \pi(2-\alpha)}{4\alpha}) & 0\cr 
         0  & \coth(\frac{\kappa \pi(2-\alpha)}{4\alpha}) }
\ee

The sliver is limit $\alpha \rightarrow 0$

\be
\tilde L_\kappa=\coth\left(\frac{\kappa \pi}{4}\right)
    \pmatrix{1 & 0\cr 
         0  & 1 }={2 \over \theta}\pmatrix{1 & 0\cr 
         0  & 1 }
\ee

We want to derive the diagonal oscillator form for the generalized butterflies.
We put

\be
n={2-\alpha \over \alpha}
\ee

\be
N=n+1={2 \over \alpha}
\ee

and we write down

\be
{Q_x\over 1+Q_x}-{1\over2} = -{1\over2}\coth\left({k\pi \over 4}\right)\tanh\left({k\pi n \over 4}\right) 
\ee

\be
{Q_y\over 1+Q_y}-{1\over2} = -{1\over2}\coth\left({k\pi \over 4}\right)\coth\left({k\pi n \over 4}\right) 
\ee

It is useful to define

\be
t=e^{-{k\pi \over 2}}
\ee

in order to simplify calculations:

\be
\coth\left({k\pi \over 4}\right)={1+t \over 1-t}
\ee

\be 
\coth\left({k\pi n \over 4}\right)={1+t^n \over 1-t^n}
\ee

\be 
\tanh\left({k\pi n \over 4}\right)={1-t^n \over 1+t^n} 
\ee

Easily we get

\be
Q_x={-t+t^n \over 1-t^{n+1}}
\ee

\be
Q_y={-t-t^n \over 1+t^{n+1}}
\ee

so

\be
|\B_g\rangle=\N_{\B_g} \, e^{-{1\over2} \,\left ( {-t+t^n \over 1-t^{n+1}} \, e^{\dagger^2}+{-t-t^n \over 1+t^{n+1}} \, o^{\dagger^2} \right)}|0\rangle
\ee

Note that for $n=1$ or alternatevely, $\alpha=1$, we find the simple Butterfly,as it must be.

\be
|\B_g (n=1)\rangle=\N_{\B} \, e^{-{1\over2} \,{2s \over 1+s^2} \, o^{\dagger^2}}|0\rangle
\ee

For our calculation in the next section we will use the diagonal basis used in \cite{fuchs}, which is the basis of eigenvectors of the operator $K_1^2$ instead of the usual $K_1$. This allows us, for instance, to write the sliver matrix in a simple two times two matrix form (see \cite{fuchs} for details)

\be\label{Sfuchs}
S= -s(k)\pmatrix{0 & 1\cr 
         1  & 0 }
\ee

where

\be
s(k)=-e^{-{k\pi \over 2}}
\ee

is the Sliver eigenvalue.
In the same basis the twist matrix $C$ is

\be
C=\pmatrix{0 & -1\cr 
         -1  & 0 }
\ee

and thus $T=CS$ becomes

\be\label{Tfuchs}
T= s(k)\pmatrix{1 & 0\cr 
         0  & 1 }
\ee

In this representation the matrix $V_{\EB}$ of the simple Butterfly was written as

\be
V_{\EB} = \frac{1}{2ch(k\pi/2)}\pmatrix{1 & 1\cr 
         1  & 1 }
\ee

Rewriting it in terms of (\ref{Sfuchs,Tfuchs})

\be
V_B={S-T\over 1-ST}
\ee

thus

\be
|\B\rangle=\N_{\B} \, e^{-{1\over2} \,{S-T\over 1-ST} \, a^{\dagger^2}}|0\rangle
\ee

which is immediately understood to be twist odd: the matrix elements $V_{nm}$ with $n+m$ even are zero while the odd ones are not, as it must be and according also with the diagonal form of such state

\be
|\B_g (n=1)\rangle=\N_{\B} \, e^{-{1\over2} \,{2s \over 1+s^2} \, o^{\dagger^2}}|0\rangle
\ee

\section{The General squeezed state form}

In this section we will show that all the string fields and star-algebra projectors we described in the previous section, can be written in a general squeezed state form involving a matrix $U$:\footnote{Kawano and Okuyama wrote down the same state with $u$ beeing a number in \cite{kawaoku}: in particular they showed that this state interpolates between the Sliver ($u=0$) and the Identity ($u=1$). In a certain sense we generalize their idea in the ``twisted'' sector of the star-algebra which is not abelian as the wedges.}

\be
|G_U\rangle={\N_\Xi \over \left (det\left(1+TU\right)\right )^{1/2}} \, e^{-{1 \over 2}\,CG\,a^\dagger a^\dagger}|0\rangle
\ee

where

\be
G=\frac{U + T}{ {\bf 1}+ TU}\
\ee

The interesting point here is that in order to obtain all relevant squeezed states studied in the star algebra, it is enough to choose the matrix $U$ in a very simple group of possibilities: the null matrix, the identity and powers of the sliver matrix and its twisted version.

The G-matrix for the ``General'' state then takes the form
 
\be\label{Gdiag}
G=\frac{U + s(k){\bf 1}}{{\bf 1} + s(k) U}\
\ee

We will refer to \cite{fuchs} for other equivalent representation of star projectors we will use in the following steps.

\subsection{$U=0$ or the Sliver}

Choosing $U=0$ in \ref{Gdiag} it is immediately to see

\be
G=s(k){\bf 1}=T
\ee

Since there is a $C$ matrix multiplication between the diagonal oscillator basis and the basis of the $a$ oscillators, we will write

\be
|G_{U=0}\rangle=\N_\Xi \, e^{-{1 \over 2}\,S\,a^\dagger a^\dagger}|0\rangle=|\Xi\rangle
\ee

which is the Sliver with the correct constant in front of the exponential factor.

\subsection{$U=1$ or the Identity}

Putting $U=1$, we have

\be
G={\bf 1}
\ee

then

\be
|G_{U=1}\rangle={\N_\Xi \over det(1+T)}\,e^{-{1 \over 2}\,C\,a^\dagger a^\dagger}|0\rangle=|I\rangle
\ee

which has also the correct constant factor. From now on, we will skip such constant factor.

\subsection{$U=C$ or the Nothing}

If we take for the matrix $U=C$, 

\be
G={C+T \over 1+S}=C
\ee

and recalling the oscillator representation of the Nothing state

\be
|G_{U=C}\rangle=e^{-{1 \over 2}\,{\bf 1}\,a^\dagger a^\dagger}|0\rangle=|\EN\rangle
\ee

\subsection{$U=-S$ or the Butterfly}

Choosing $U=-S$ and using for $S$ the form

\be\label{Sfuchs}
S= -s(k)\pmatrix{0 & 1\cr 
         1  & 0 }
\ee

we have

\be
G={s(k)\pmatrix{0 & 1\cr 
         1  & 0 } +s(k)\pmatrix{1 & 0\cr 
         0  & 1 } \over \pmatrix{1 & 0\cr 
         0  & 1 }+s(k)s(k)\pmatrix{0 & 1\cr 
         1  & 0 } }
\ee

then 

\be
G={s(k) \over 1 + s^2(k)} \pmatrix{1 & 1\cr 
         1  & 1 }=-\frac{1}{2ch(k\pi/2)}\pmatrix{1 & 1\cr 
         1  & 1 }
\ee

which is the diagonal $K_1^2$ representation of the butterfly, as showed in \cite{fuchs}. Thus

\be
|G_{U=-S}\rangle=e^{-{1 \over 2}\,V_{\EB} a^\dagger a^\dagger}|0\rangle=|\EB\rangle
\ee

\subsection{$U=(-T)^{N-1}$ or the Wedge state}

Putting $U=(-T)^{N-1}$ and recalling that $T$ is simply proportional to the identity matrix in two dimension

\be
G={\left(-s(k)\right)^{N-1}\pmatrix{1 & 0\cr 
         0  & 1 }^{N-1} +s(k)\pmatrix{1 & 0\cr 
         0  & 1 }\over \pmatrix{1 & 0\cr 
         0  & 1 } + s(k) \left(-s(k)\right)^{N-1}\pmatrix{1 & 0\cr 
         0  & 1 }^N} 
\ee

which gives

\be
G={\left(-s(k)\right)^{N-1} +s(k) \over 1 - \left(-s(k)\right)^N}\pmatrix{1 & 0\cr 
         0  & 1 }
\ee

which is exactly the definition of the Nth wedge state matrix $T_N$, recalling that in our basis.
So, we get

\be
|G_{U=(-T)^{N-1}}\rangle=e^{-{1 \over 2}\,CT_N\,a^\dagger a^\dagger}|0\rangle=|\ET_N\rangle
\ee

\subsection{$U=(-S)^{2N-1}$ or the Generalized Butterfly}

It is important to notice that in this formalism, apart the constant s(k), the matrix $T$ behaves as an even power of the matrix $S$. In this sense, even powers of matrix $S$ lead to wedge states, odd ones to generalized butterflies as we will see now. Indeed, taking $U=(-S)^{N-1}$ with $N$ even, we get
 
\be
G={\left(-s(k)\right)^{N-1}\pmatrix{0 & 1\cr 
         1  & 0 }^{N-1} +s(k)\pmatrix{1 & 0\cr 
         0  & 1 }\over \pmatrix{1 & 0\cr 
         0  & 1 } + \left(s(k)\right)^N\pmatrix{0 & 1\cr 
         1  & 0 }^{N-1}} 
\ee

with $N$ even. After a bit of algebra

\begin{eqnarray}
G=\frac{ \pmatrix{s & s^{N-1}\cr 
         s^{N-1}  & s }}{ \pmatrix{1 & s^N\cr 
         s^N  & 1 }}
 =  {s \over 1-s^{2N}}\pmatrix{1-s^{2N-2} & s^{N-2}(1-s^2)\cr 
         s^{N-2}(1-s^2)  & 1-s^{2N-2} }
\end{eqnarray}

which can be rewritten as

\be
G={s \over 1-s^{2N}}\left[(1-s^{2N-2})\I - s^{N-2}(1-s^2)C\right]
\ee

Let us separate the twist even and twist odd sectors

\be
G_{even}={s \over 1-s^{2N}}\left[(1-s^{2N-2}) - s^{N-2}(1-s^2)\right]
\ee

\be
G_{odd}={s \over 1-s^{2N}}\left[(1-s^{2N-2}) + s^{N-2}(1-s^2)\right]
\ee

in order to write

\be
|G\rangle=\, e^{-{1\over2} \,\left ( G_{even} \, e^{\dagger^2}+G_{odd}\, o^{\dagger^2} \right)}|0\rangle
\ee

with

\be
G_{even}={s - s^{N-1} \over 1-s^{N}}
\ee

and

\be
G_{odd}={s + s^{N-1} \over 1 + s^{N}}
\ee

In order to compare this with the expected state, we need the diagonal oscillator representation of generalized butterflies we derived in the previous section

\be
|\B_g\rangle=\N_{\B_g} \, e^{-{1\over2} \,\left ( {-t+t^n \over 1-t^{n+1}} \, e^{\dagger^2}+{-t-t^n \over 1+t^{n+1}} \, o^{\dagger^2} \right)}|0\rangle
\ee

Now, we compare the form for the Generalized Butterfly we found with the General state with $U=\left(-S\right )^{N-1}$, $N$ even

\be
|G_{U=\left(-S\right)^{N-1}}\rangle=\, e^{-{1\over2} \,\left ( {s - s^{N-1} \over 1-s^{N}} \, e^{\dagger^2}+ {s + s^{N-1} \over 1 + s^{N}}\, o^{\dagger^2} \right)}|0\rangle
\ee

Recalling $-t=s$, $N=n+1$, it is straightforward to see that the even sector  of $|\EB_g\rangle$

\be
{-t+t^n \over 1-t^{n+1}}={s+(-s)^{N-1} \over 1-(-s)^N}={s - s^{N-1} \over 1 - s^N}
\ee

coincides with $G_{even}$ and the odd sectors of  $|\EB_g\rangle$

\be
{-t - t^n \over 1 + t^{n+1}}={s + s^{N-1} \over 1 + s^N}
\ee

coincides with  $G_{odd}$.

\section{General state from Laguerre states}

Kawano and Okuyama showed in \cite{kawaoku} that if $P_n$ are projectors and form a complete basis, the function 

\be
P(u)=\sum_{n=0}^{\infty}u^n P_n
\ee

with $u$ real constant, can be written as

\be
|P(u)\rangle={1 \over \left (det\left(1+Tu\right) \right)^{1/2}} \, e^{-{1 \over 2}\,\frac{u + CS}{ {\bf 1}+ Tu}\,a^\dagger a^\dagger}|0\rangle
\ee

and has the property

\be
P(u)*P(v)=P(uv)
\ee

under star product, as can be checked using the formula \cite{kawaoku}

\be
e^{\frac 12 a^{\dagger} CT_1 a^{\dagger}}|0\rangle \ast e^{\frac 12 a^{\dagger} CT_2 a^{\dagger}}|0\rangle = det^{-\frac 12}\left[1-X(T_1+T_2)+XT_1T_2 \right] e^{\frac 12 a^{\dagger} CT_{12} a^{\dagger}}|0\rangle 
\ee

with

\be
T_{12} = \frac{X-X(T_1+T_2)+T_1T_2}{1-X(T_1+T_2)+XT_1T_2}
\ee

obtained assumig that  $C$, $T_1$ , $T_2$ and $X$ commute with each other.

$P(U)$ interpolates between the Sliver ($u=0$) and the Identity state ($u=1$). Indeed,

\be
P(u=1)=\sum_{n=0}^{\infty} P_n =|I\rangle
\ee

In \cite{BMS3} was defined an infinite class of star projectors obtained acting on the Sliver with a Laguerre polynomial expression of quadratic creation operators. They were called ``Ancestors'' because in the low energy limit give rise to all GMS solitons. In this section we will use a modified form of such Ancestors as $P_n$: we will call them ``Laguerre'' states and following \cite{kawaoku} we will define a similar function with a generalization about the constant $u$: we will use a matrix $U$. We will see that the Laguerre form of such states will lead to the General Squeezed state we saw in the previous section interpolating between Wedges and Butterflies.  
Let us recall the definition and properties of Ancestors

\be
|\Lambda_n\rangle = (-\kappa)^n L_n\Big(\frac{\x}{\kappa}\Big) |\Xi\rangle
\label{Lambdan}
\ee

where

\be
\x = (a^\dagger  \xi)\, (a^\dagger C' \zeta)   \label{x}
\ee

and the two vectors inside $\xi=\{\xi^\mu_{n}\}$ and $\zeta =
\{\zeta^\mu_{n}\}$ are chosen to satisfy the conditions

\be
\rho_1 \xi =0,\quad\quad \rho_2 \xi =\xi, \quad\quad
{\rm and}\quad \rho_1 \zeta =0,\quad\quad \rho_2\zeta=\zeta,
\label{xizeta}
\ee

 and

\be
\xi^T \frac 1{1-T^2}\zeta =-1 ,\quad\quad
\xi^T \frac {T}{1-T^2}\zeta= -\kappa
\label{cond0}
\ee

and $\kappa$ is an arbitrary real constant. $|\Xi\rangle$ is the sliver. 

Hermiticity requires that
\be
(a \xi^*)(aC \zeta^*) = (aC\xi)(a \zeta)\label{hermit}
\ee

We know that
\begin{eqnarray}
 |\Lambda_n\rangle * |\Lambda_m\rangle = \delta_{n,m}
|\Lambda_n\rangle \label{nstarm}\\
 \langle \Lambda_n |\Lambda_m\rangle = \delta_{n,m}
 \langle \Lambda_0 |\Lambda_0\rangle \label{nm}
\end{eqnarray}

Now we define Laguerre states $|\widehat \Lambda_n \rangle$ as

\be
|\widehat \Lambda_n =  (-\widehat \kappa)^n L_n\Big(\frac{\widehat \x}{\widehat \kappa}\Big) |\Xi\rangle
\ee

where $\widehat \kappa$ is chosen to be 

\be
\kappa = Tr(T)
\ee

and $\widehat \x$ is defined 
\be
\x = -{1 \over 2}\left(C \left(1-S^2\right)\right)_{mn}a^{\dagger}_m a^{\dagger}_n
\ee

In the diagonal basis these assumptions, omitting the integral over the continous parameter k, become
 
\be
\kappa = s(k)
\ee

\be
\widehat \x = -{1 \over 2}\left(1-s^2(k)\right)\left( e^{\dagger^2}+ o^{\dagger^2}\right)
\ee

After this modification respect to Ancestors $|\Lambda_n\rangle$, we cannot say they are still projectors because it is not trivial to check conditions (\ref{xizeta}) and (\ref{cond0}). In any case, we do not need Laguerre states $|\widehat \Lambda_n\rangle$ to be projectors: we use their explicit form.
If we define the function

\be
G(U) = \sum_{n=0}^{\infty}U^n|\widehat \Lambda_n\rangle
\ee

using a well--known resummation formula for Laguerre polynomials 

\be
\sum_{n=0}^{\infty}z^n L_n(x)
= \frac 1{1-z}e^{-\frac {zx}{1-z}}
\ee

and putting

\be
z = -s(k)U
\ee

\be
\widehat \x = -{1 \over 2}\left(1-s^2(k)\right)\left( e^{\dagger^2}+ o^{\dagger^2}\right)
\ee

and recalling the argument of Laguerre polynomial is $\frac{\widehat \x}{\widehat \kappa}$, we obtain

\be
G(U) = \frac {\N_\Xi}{\left(det(1+s(k)U)\right)^{1/2}}e^{\frac {U}{1+s(k)U}\left[-{1 \over 2}\left(1-s^2(k)\right)\left( e^{\dagger^2}+ o^{\dagger^2}\right)\right]-{1 \over 2}s(k)\left( e^{\dagger^2}+ o^{\dagger^2}\right)}|0\rangle
\ee

and finally

\be
G(U) = {\N_\Xi \over \left(det\left(1+TU\right)\right)^{1/2}} \, e^{-{1 \over 2}\,\frac{U + T}{ {\bf 1}+ TU}\,\left( e^{\dagger^2}+ o^{\dagger^2}\right)}|0\rangle
\ee

which coincides with the General Squeezed state $|G_U\rangle = G(U)$.\\

\medskip

Although it is far to be proved that $|\widehat \Lambda_n \rangle$ are still projectors, they seem to behave as a basis for surface states.
In particular, we notice 

\be
|G(U=1)\rangle=\sum_{n=0}^{\infty}|\widehat \Lambda_n\rangle = |I\rangle
\ee

In the next section we will remark that looking at the partial isometry symmetry of SFT, also the sum of the $|\Lambda_n$ which are certainly projectors behaves in some sense as a sort of Identity, perhaps of a star subalgebra.

\section{Partial Isometry symmetry}

Let us recall the general definition of partial isometry.
Given a lagrangian invariant under a group of unitary transformations on Hilbert space $U(H)$, then 

\be
\phi \rightarrow U \phi U^{\dagger}
\ee

with

\be
UU^{\dagger}=U^{\dagger}U=1
\ee

leaves the action invariant and takes solutions of the equation of motion to solutions since

\be
\frac{\d V}{\d \phi}\rightarrow U\left(\frac{\d V}{\d \phi}\right)U^{\dagger}
\ee

But, in order to show that solutions transform into solutions, it is only necessary to use $U^{\dagger}U=1$, since it is required that fields transform covariantly.Operators U such that are called isometries because they preserve metric on Hilbert space:

\be
\langle\chi|\psi\rangle \rightarrow \langle\chi|U^{\dagger}U|\psi\rangle=\langle\chi|\psi\rangle
\ee

If it is also true that $UU^{\dagger}=1$ then $U$ is unitary. In a finite dimensional Hilbert space it is always true but not in a infinite dimensional one.
Thus if we find a nonunitary isometry it will still map solutions to solutions, but these solutions will not be related by the global symmetry (or gauge symmetry if we add gauge fields) of the action. So, you will obtain new solutions.
A typical example of nonunitary isometry is the shift operator
$S: |n\rangle \rightarrow |n+1\rangle$ :

\be
S = \sum_{n=0}^{\infty} | n+1 \rangle\langle n|
\ee

Obviously, 

\be
S^{\dagger}S=1 
\ee

but 

\be
SS^{\dagger}=1-P=1-|0\rangle\langle0|
\ee

More generally, $U=S^n$ is a nonunitary isometry and 

\be
UU^{\dagger}=1-P_n
\ee

with

\be
P_n=\sum_{k=0}^{n-1}|k\rangle\langle k|
\ee

Partial isometry can be applied in a solution generating technique in noncommutative field theories (see for instance \cite{Komaba} and reference therein): we can start with a trivial constant solution $\phi=\lambda_iI$ with $I$ the identity operator and $\lambda$ a global miminum for the potential and transforming with $U=S^n$ we obtain the new solution

\be
\phi=S^n\lambda_iIS^{\dagger^n}=\lambda_i\left(1-P_n\right)
\ee

This solution, for instance, will describe a finite energy excitation above the vacuum.
A similar structure arise also in VSFT. In fact we can 
easily construct the correspondents of the operators $|n\rangle\langle m|$ \cite{tope} and then of the shift operator.\\
Let us first define  

\be
X= a^\dagger \tau \xi \, \quad\quad Y = a^\dagger C'\zeta\label{XY}
\ee

so that $\x = XY$. The definitions we are looking for are as follows

\be 
|\Lambda_{n,m}\rangle \, = \, \sqrt \frac {n!}{m!} (-\kappa)^n \,
Y^{m-n} L_n^{m-n}
\left(\frac \x \kappa\right)|\ES_\perp\rangle, 
\quad\quad n\leq m \label{nm1}
\ee

\be
|\Lambda_{n,m}\rangle \, = \, \sqrt \frac {m!}{n!} (-\kappa)^m \, 
X^{n-m} L_m^{n-m}
\left(\frac \x \kappa\right)|\ES_\perp\rangle, 
\quad\quad n\geq m\label{nm2}
\ee

where $L_n^{m-n}(z) = \sum_{k=0}^m \left(\matrix {m\cr n-k}\right) (-z)^k/k!$.
With the same techniques as in \cite{BMS3} one can prove that

\be
|\Lambda_{n,m}\rangle * |\Lambda_{r,s}\rangle = \delta_{m,r}|\Lambda_{n,s}\rangle  
\label{LL}
\ee

for all natural numbers $n,m,r,s$. It is clear that the previous states 
$|\Lambda_n\rangle$ coincide with $|\Lambda_{n,n}\rangle$. Therefore, following \cite{GMS}, \cite{Komaba}, 
we can apply to the construction of projectors in the VSFT star algebra the
solution generating technique, in the same way as in the harmonic oscillator
Hilbert space ${\cal H}$.
We write the analog of the shift operator in VSFT as

\be
S = \sum_{n=0}^{\infty}|\Lambda_{n+1,n}\rangle
\ee
\be
S^{\dagger} = \sum_{m=0}^{\infty}|\Lambda_{m,m+1}\rangle
\ee
Then follows

\be
S*S^{\dagger} = \sum_{n,m=0}^{\infty}|\Lambda_{n+1,n}\rangle * |\Lambda_{m,m+1}\rangle = \sum_{n=0}^{\infty}|\Lambda_{n}\rangle - |\Lambda_{0}\rangle
\ee
\be
S^{\dagger}*S = \sum_{n,m=0}^{\infty}|\Lambda_{n,n+1}\rangle * |\Lambda_{m+1,m}\rangle  = \sum_{n=0}^{\infty}|\Lambda_{n}\rangle 
\ee

which looks like the partial isometry relations 

\be
SS^{\dagger}=1-P=1-|0\rangle\langle0|
\ee
\be
S^{\dagger}S=1 
\ee

and it works among Ancestors as the solution generating technique we recalled \cite{Komaba} since

\be
\Lambda_{n+m}=S^m \Lambda_n S^{\dagger^m}
\ee

Thus, at least among such solutions, the sum of Ancestors takes the role of the identity string field.

\section{Conclusions}
Wedge States are surface states interpolating between the Sliver and the Identity string field. Butterflies interpolate between Sliver and the ``Nothing'' state. In this paper we have showed a compact way to write the Butterflies in an oscillator form using powers of the Sliver $S$ matrix and its twist version $T$ as was done for the Wedges in terms of the matrix $T$. We have also written a squeezed state depending on a matrix $U$, whose form give rise to Butterflies or Wedges depending on the choice of powers of $S$ or $T$ matrices as $U$. In the literature, the definition of Wedges and Butterflies involves, respectevely, an integer number $N$ and a real parameter $0 \leq \alpha \leq 2$. The identity is seen as the $N=1$ wedge state, the wedge of zero angle, the Sliver as the $\infty$-wedge, the wedge of infinite angle. On the other hand, the Sliver is also seen as the $\alpha \to 0$ limit of Butterflies and the Nothing is the $\alpha=2$ case. To be more precise, in this paper we have studied only the Butterflies whose $\alpha$-paramenter can be written as $\alpha= \frac 2N$. In this view, the interpolating state starts from the Nothing ($U=C$) and growing in the powers of the $S$ matrix ($U=(-S)^{2N-1}$) reachs the Sliver at infinity. Alternatevely, it starts from the twist version of the Nothing which is the Identity ($U=1$) and growing with powers of twisted version of $S$ matrix ($U=(-T)^{N-1}$) becomes the Sliver at infinity.

\begin{figure}
\begin{center}
\leavevmode
\includegraphics[scale=0.5]{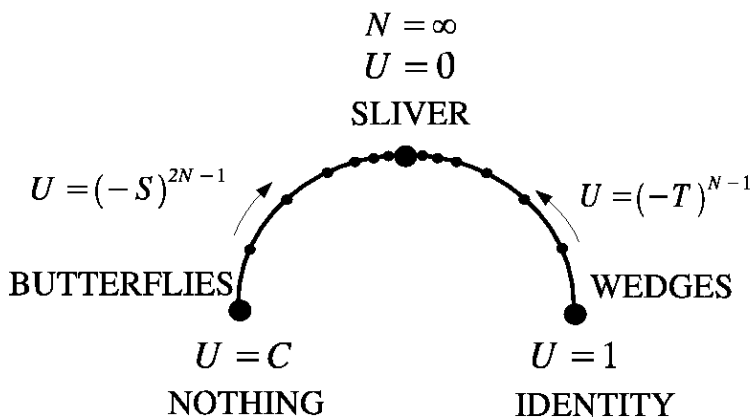}
\caption{Interpolation between the Nothing state and the Sliver and between the Identity and the Sliver}\label{dise2}
\end{center}
\end{figure}

Taking inspiration from star projectors found in \cite{BMS3}, we defined a class of states $|\widehat \Lambda_n$, that we cannot say at the moment if they are or not projectors and we called Laguerre States. Following the idea in \cite{kawaoku} but using explicit form of Laguerre states instead of the fact they are projectors, we gave a formal derivation of such General Squeezed State. 
Posponing the study of such Laguerre states in a subsequent paper, we notice that if they would be projectors

\be
|\widehat \Lambda_n \rangle \ast |\widehat \Lambda_m \rangle= \delta_{mn} |\widehat \Lambda_n \rangle
\ee

 although the derivation of the interpolating state is rather formal since every state of the resummation formula is a squeezed state times a matrix, it would be possible to see the multiplication rule between wedge states because we would have:

\be
|\ET_N\rangle \ast |\ET_M\rangle = \sum_{n=0}^{\infty} \left ((-T)^{N-1}\right )^n |\widehat \Lambda_n \rangle \ast \sum_{m=0}^{\infty} \left ((-T)^{M-1}\right )^m |\widehat \Lambda_m \rangle = \sum_{n=0}^{\infty} \left ((-T)^{M+N-1-1}\right )^n |\widehat \Lambda_n \rangle
\ee

thus

\be
|\ET_N\rangle \ast |\ET_M\rangle = |\ET_{M+N-1}\rangle
\ee   
   
In any case, despite Laguerre states are or not star projectors, the same check for Butterflies using oscillator formalism is complicated by the fact $S$ matrix does not commute with $X = CV^{11}$ matrix of the Witten's vertex.

\medskip

\noindent {\bf Acknoledgements}

I would like to thank Loriano Bonora for useful comments and discussions. I would like to thank also Luis F. Alday, Carlo Maccaferri and Serguey Petcov.

\end{document}